\documentclass[pra,twocolumn]{revtex4}

\usepackage{amsmath}
\usepackage{amssymb}
\usepackage{graphicx}
\usepackage{booktabs}
\usepackage{subfig}
\usepackage{bm}

\newcommand{\units}[2]{#1\textrm{\thinspace #2}}

\newcommand{\zt}{y(t)}
\newcommand{\xt}{x(t)}
\newcommand{\wt}{z(t)}
\newcommand{\zk}{y_m}
\newcommand{\Zz}{Y}
\newcommand{\zz}{y}
\newcommand{\xk}{s_m}
\newcommand{\Xx}{S}

\newcommand{\sk}{x_m}
\newcommand{\Ss}{X}
\newcommand{\sss}{x}
\newcommand{\wk}{z_m}
\newcommand{\Ww}{Z}

\newcommand{\zkp}{y_m'}

\newcommand{\pd}{p}
\DeclareMathOperator{\Var}{Var}
\DeclareMathOperator{\E}{E}
\DeclareMathOperator{\re}{Re}

\newcommand{\capc}{\text{C}}

\newcommand{\Ew}{E_{n}}

\newcommand{\capERC}{\text{C}}
\newcommand{\capClas}{\capc_\text{Clas}}
\newcommand{\capQCoh}{\capc_\text{Coh}}
\newcommand{\capAEN}{\capc_\text{AEN}}
\newcommand{\es}{\varepsilon_s}
\newcommand{\esStar}{\varepsilon_{s}^*}
\newcommand{\ew}{\varepsilon_w}
\newcommand{\mInf}[2]{I(#1;#2)}

\newcommand{\snr}{{\rm{SNR}}}

\graphicspath{{../../Pictures/}}

\begin{document}

\title{Information Rate Loss due to Radiation Decoherence}

\author{Alfonso Martinez}
\affiliation{Department of Electrical Engineering, Technische Universiteit Eindhoven, Eindhoven, The Netherlands}
\date{\today}

\begin{abstract}
The information rates achievable by using electromagnetic radiation affected by thermal noise and signal decoherence are studied. The standard coherent Gaussian model is compared with an alternative photon gas model which represents lack of
a shared phase reference between transmitter and receiver. 
At any frequency, information rates over the photon gas model essentially coincide with those over the Gaussian model when the signal-to-noise ratio is below a threshold. Only above the threshold does decoherence cause a loss in information rates; the loss can amount to half of the capacity. The threshold exceeds \units{40}{dB} for radio frequencies and vanishes at optical frequencies. 
\end{abstract}

\maketitle

\section{Introduction}

The consideration of quantum effects in information theory has a rich history, starting with the pioneering work of Gordon \cite{gordon64:noiseOpticalFrequencies}, and later pursued, among others, by Helstrom \cite{helstrom76:quantumDetection} and Holevo \cite{holevo99:capacityQuantumGaussian,holevo01:evaluatingCapacityBosonicGaussian}. A trait shared by these works is the special role played by thermal noise and coherent states, the natural quantum counterpart of the Gaussian model from classical communication theory.


Inspired by recent work on reference frames in information theory \cite{bartlett07:referenceFramesSuperselectionRules}, where Schumacher is quoted as saying that ``restrictions on the resources available for communication yield interesting communication theories'', we consider, in addition to thermal noise, the effect of decoherence, by which we mean absence of a shared phase reference between transmitter and receiver, and deal with the information rate loss incurred by such restriction, viz.\ decoherence between transmitter and receiver.

Our work is related to the analysis of direct detection methods at optical frequencies \cite{caves94:quantumLimitsBosonic} in the sense that communication is non-coherent. However, the discrete channel model we use is different, as it represents the radiation field as a photon gas. Details are given in Sec.~\ref{sec:semiclassicalModel} from a semiclassical perspective.
The key element is that information is sent by modulating the energy of the Fourier modes of the field; at the receiver, energy is measured.

In Sec.~\ref{sec:channelCapacity} we determine the channel capacity of the photon gas model and derive the main result of this paper, namely that decoherence leads to no loss in information rate when the signal-to-noise ratio lies below a threshold; above the threshold, up to half of the capacity is lost.
For a frequency $\nu$ (in Hertz), this threshold is approximately given by
  $\frac{6\cdot 10^{12}}{\nu}$
at \units{290}{K}, 
and is thus large only for radio and microwave frequencies. 

Previous studies of direct detection \cite{katz04:capacityAchievingNoncoherentAWGNChannels} showed a non-negligible capacity penalty. In Sec.~\ref{sec:additiveEnergyField} we relate this discrepancy to a different way of accounting for the energy of a mixture of thermal and coherent radiation.


\section{Semiclassical Analysis}
\label{sec:semiclassicalModel}
In this section, we present a semiclassical analysis of the effect of radiation decoherence.
We consider one polarization of the electromagnetic field at an aperture, which we denote by $\zt$, a complex-valued function. As is well-known,
%
we can perform a Fourier decomposition of the function $\zt$ onto
frequencies of the form $\nu_c+\frac{m}{T}$, lying in a band of width $W$ around a reference frequency $\nu_c$; here $T$ is the duration of the observation interval.

Further, the function $\zt$
is the sum $\xt + \wt$, where the signal $\xt$ and noise $\wt$ are respectively given by 
\begin{align}
    \xt &= \sum_m \tfrac{1}{\sqrt{T}}\sk e^{i2\pi (\nu_c+\frac{m}{T})t} \label{eq:xtCoh}\\
    \wt &= \sum_m \tfrac{1}{\sqrt{T}}\wk e^{i2\pi (\nu_c+\frac{m}{T})t};
\end{align}
here $\sk$ is the useful signal of mode $m$, set at the transmitter (except for a propagation loss and a phase rotation), and $\wk$ are samples of Gaussian noise; all quantities are complex. 
With coherent detection, the quantities $\zk = \xk + \wk$ are recovered at the receiver \cite{caves94:quantumLimitsBosonic}.

\subsection{Radiation Decoherence and Photon Gas Model}

An implicit assumption behind Eq.~\eqref{eq:xtCoh} is the existence of a shared phase reference between transmitter and receiver. In general, however, decoherence takes place and a phase drift $\phi_m(t)$, possibly a function of the Fourier mode $m$, appears. Following Lax's analysis of oscillator noise \cite{lax67:classicalNoiseV}, we model the phase drift $\phi_m(t)$ as a Brownian motion with zero mean and variance $\E[\phi_m^2(t)] = 2\pi\beta_m t$.
The total signal at the receiving aperture is still given by $\zt = \xt + \wt$, but the transmitted signal is now
\begin{equation}
    \xt = \sum_m \tfrac{1}{\sqrt{T}}\sk e^{i\phi_m(t)}e^{i 2\pi (\nu_c+\frac{m}{T})t}.
\end{equation}

When $\beta_m = 0$, we obviously recover the coherent model. However, for any $\beta_m > 0$, however tiny, as the observation duration increases $T\to\infty$, then the signal in the $m$-th output of a coherent receiver vanishes, i.\ e.\
\begin{equation}\label{eq:xkNon-Coherent}
    \tfrac{1}{\sqrt{T}}\int_0^T \sss(t)e^{-i 2\pi (\nu_c+\frac{m}{T})t}\,dt = \tfrac{\sk}{T}\int_0^T e^{i\phi_m(t)}\,dt\simeq 0.
\end{equation}
In this case, a different detection method is required.
Fortunately, the energy in the $m$-th mode is well defined (see Sec.~\ref{sec:additiveEnergyField}) by the sum $|\sk|^2 + |\wk|^2$. This suggests using a form of direct detection, which we now describe.

By construction the signal $\xt$ is a mixture of independent frequency tones.
Demultiplexing them at the receiver generates a set of parallel signals $\zeta'_m(t)$ given by
\begin{align}\label{eq:zetaKPrime}
    \zeta_m'(t) = \tfrac{1}{\sqrt{T}}\bigl(\sk e^{i\phi_m(t)}+\wk\bigr)e^{i 2\pi (\nu_c+\frac{m}{T})t}.
\end{align}
The instantaneous energy $\zeta_m''(t) = |\zeta_m'(t)|^2$ can be integrated in the interval
$(0,T)$ to generate an output $\zkp$,
\begin{align}
    \zkp 
    = \tfrac{1}{T}\int \bigl(|\sk|^2 + |\wk|^2 + 2\re(\sk e^{i\phi_m(t)}\wk^*)\bigr)\,dt.
\end{align}

There are now two possibilities, depending on how fast the phase noise $\phi_m(t)$ varies in time.
If $\phi_m(t)$ is constant, i.\ e.\ $\beta_m = 0$, the output $\zkp$ is given by
\begin{align}\label{eq:zkp-nonCoh}
    \zkp = |\sk e^{i\phi_m} + \wk|^2,
\end{align}
as for the squared output of a coherent receiver. In the approximation that the energy is continuous, $\zkp$ follows a non-central chi-square distribution; its square root $\sqrt{\zkp}$ is distributed according to a Rician distribution.
On the other hand, if the energy is discrete, the distribution of $\zkp$ is Laguerre with parameters $|\sk|^2$ and $|\wk|^2$ \cite{haus62:quantumNoiseLinearAmplif,glauber63:_coher_incoh_states_radiat_field}.

In the second alternative, we take $\beta_m T \to\infty$, and then
\begin{align}\label{eq:zk-dd}
    \zkp &\simeq \tfrac{1}{T}\int \bigl(|\sk|^2 + |\wk|^2\bigr)\,dt = |\sk|^2 + |\wk|^2,
\end{align}
namely the sum of the energies of signal and noise. As mentioned previously, this condition holds true as $T\to\infty$ as long as $\beta_m > 0$. If the energy is assumed discrete, then the signal component is modelled as a Poisson random variable and the additive noise has a Bose-Einstein distribution \cite{caves94:quantumLimitsBosonic}. One can think of this model as a photon gas, where the receiver counts the number of photons in each Fourier mode.
Otherwise, for a continuous-energy approximation, the noise energy $|\wk|^2$ has an exponential density, which is both the limiting form of a Bose-Einstein distribution and the density of the squared amplitude of complex Gaussian noise \cite{feller71:introductionProbability2}.

The output $\zkp$ in Eq.~\eqref{eq:zk-dd} is the sum of the energies of signal and noise.
It is worthwhile noting that this additivity in energy does not hold for the Laguerre distribution, the standard result for the superposition of a thermal field and a coherent state. In the next section we briefly review the derivation of the Laguerre distribution and discuss the conditions under which energy is additive. Then, in Sec.~\ref{sec:channelCapacity} we compare the information rates over the Gaussian model with those achievable by an energy measurement. We then use the difference in information rates between the coherent and non-coherent models as an estimate of the effect of decoherence.

\subsection{Additive Energy versus Additive Field}
\label{sec:additiveEnergyField}

For simplicity, let us consider the superposition of two continuous-time signals $x_1(t)$ and $x_2(t)$. The instantaneous energy $\zeta''(t)$ is given by
\begin{align}
    \zeta''(t) = \bigl|x_1(t)\bigr|^2 + \bigl|x_2(t)\bigr|^2 + 2\re\bigl(x_1(t)x_2^*(t)\bigr),
\end{align}
with a beat term $\re\bigl(x_1(t) x_2^*(t)\bigr)$. Even though its mean is zero, it is not identically zero.
Moreover it makes the total energy after integration different from the sum of the energies of $x_1(t)$ and $x_2(t)$, as in Eq.~\eqref{eq:zkp-nonCoh}.

We solve this issue by noting that the superposition model is not unitary and choose instead a unitary matrix to represent it. Consider such a matrix $U$,
\begin{equation}
    U = \frac{1}{\sqrt{2}}\begin{pmatrix} 1 & e^{i\phi}\\ -e^{i\phi} & 1\end{pmatrix},
\end{equation}
where $\phi$ is a phase offset. Instead of $x_1(t) + x_2(t)$, there are now two outputs, say $y_1(t)$ and $y_2(t)$, linear combinations of the inputs, namely
\begin{align}
    y_1(t) &= \frac{1}{\sqrt{2}} \bigl(x_1(t) + e^{i\phi} x_2(t)\bigr) \\
    y_2(t) &= \frac{1}{\sqrt{2}} \bigl(-e^{i\phi}x_1(t) + x_2(t)\bigr).
\end{align}
The instantaneous energy at each output, $\zeta_{1,2}''(t)$, is then
\begin{align}
    \zeta_{1,2}''(t) &= \frac{1}{2} \Bigl(\bigl|x_1(t)\bigr|^2 + \bigl|x_2(t)\bigr|^2 \pm 2\re\bigl(x_1(t)e^{i\phi}x_2^*(t)\bigr)\Bigr),
\end{align}
where the sign $+(-)$ goes with output $1(2)$.
The total output is the sum over the two branches, $\zeta_1''(t) + \zeta_2''(t)$.

Each output, observed individually, follows a Laguerre distribution. However, the outputs are not independent but correlated through the beat term. As these terms have different sign, they cancel out when we sum the two outputs, and we indeed see that the total energy is the sum of the energy in the signal and the noise components.

\section{Information Rates}
\label{sec:channelCapacity}

\subsection{Discrete Models for Radiation}

We have discussed two models, viz.\ coherent detection and energy detection. For the latter, we have in turn described two options, a photon gas model where the energy is discrete and an exponential noise model under the approximation that the energy is continuous.
All cases are described by a channel model of the form
\begin{equation}
    \zk = \xk(\sk) + \wk, \qquad m = 1, \dotsc, n,
\end{equation}
where $\zk$ is a measurement on the $m$-th Fourier mode, $\sk$ the $m$-th signal component, $\xk$ the useful signal at the output, and $\wk$ the $m$-th sample of additive noise.

The specifics of each model are
\begin{enumerate}
\item For coherent detection, $\zk$, $\xk = \sk$, and $\wk$ are complex-valued. Further, $\wk$ are samples of  Gaussian noise with variance $\sigma^2$.
\item For the photon gas model, $\zk$, $\xk$, and $\wk$ are non-negative integers, a number of photons of energy $h\nu$ each. The signal component $\xk$ has a Poisson distribution with mean $\sk$, and the noise component $\wk$ has an Bose-Einstein (or geometric) distribution with mean $\ew = (e^\frac{h\nu}{kT}-1)^{-1}$, as thermal radiation. Here $\nu$ is assumed constant for all modes, and equal to $\nu_c$; $T$ is a temperature.
\item For continuous energy, $\zk$, $\xk = \sk$ and $\wk$ are non-negative real numbers. Then, $\wk$ are samples of exponential noise with mean $\Ew$.
\end{enumerate}
In all cases, a constraint on the signal energy $E_s = \es h\nu$ is imposed, where $\es$ is the average number of signal photons, and $E_s$ the corresponding energy. Moreover, it is convenient to set $\sigma^2 = \Ew = \ew h\nu$. For the first and third models, we define an average signal-to-noise ratio $\snr$ as $\snr = E_s/\sigma^2 = E_s/\Ew$.

%
%

The largest information rate that can be sent over a channel with output conditional density $\pd_{\Zz|\Ss}(\zz|\sss)$ is the channel capacity $\capc$ \cite{cover91:informationTheory}, given by
\begin{equation}
  \capc = \sup_{\pd_\Ss(\sss)} I(\Ss;\Zz),
\end{equation}
where the maximization is over all input densities $\pd_\Ss(\sss)$ satisfying the energy constraint, and $I(\Ss;\Zz)$ is the mutual information between channel input and output. For continuous output the mutual information is given by
\begin{equation}
  I(\Ss;\Zz) = \int \pd_{\Ss}(\sss)\int \pd_{\Zz|\Ss}(\zz|\sss)\log\frac{\pd_{\Zz|\Ss}(\zz|\sss)}{\pd_\Zz(\zz)}\,d\zz\,d\sss,
\end{equation}
where $\pd_\Zz(\zz) = \int \pd_\Ss(\sss)\pd_{\Zz|\Ss}(\zz|\sss)\,d\sss$. For discrete output, the integrals over $\zz$ should be replaced by sums.

\subsection{Capacity with Coherent Detection}

In classical information theory, the capacity $\capClas$ is given by the well-known Shannon formula \cite{cover91:informationTheory},
\begin{equation}\label{eq:capacityClassical}
    \capClas(E_s,\sigma^2) = \log\Biggl(1+\frac{E_s}{\sigma^2}\Biggr),
\end{equation}
where $\sigma^2 = kT$ is a common approximation to the noise spectral density.
In addition, with a quantum model for measurement, noise is additive Gaussian with variance $(\ew + 1)h\nu$ \cite{gordon62:quantumEffectsCommunicationSystems,caves94:quantumLimitsBosonic}, and one obtains the quantum capacity with coherent detection $\capQCoh$, given by
\begin{align}
    \capQCoh(\es,\ew) 
    &= \log\Biggl(1+\frac{\es }{\ew +1}\Biggr)\label{eq:capacityCoherentQuantum}.
\end{align}
At radio and microwave frequencies, $\ew \simeq \frac{h\nu}{kT}\gg 1$ and we recover Shannon's formula,
\begin{equation}
    \capQCoh(\es,\ew) \simeq \capClas(E_s,kT).
\end{equation}
These expressions give the information rates with thermal noise and in absence of decoherence.


%

\subsection{Capacity of the Photon Gas}

In the photon gas model, two sources of noise are present at the output: Poisson noise, arising from the signal itself, and additive noise. Distinct behaviour is to be expected depending on which noise prevails.
In a first approximation, the behaviour is determined by the noise variance. The additive noise variance is given by $\ew(1+\ew)$ (it follows a Bose-Einstein distribution), whereas the signal variance $\es$ (it is a Poisson random variable) \cite{feller71:introductionProbability2}.
Of practical importance is the region where $\ew \gg 1$, for which the variances coincide when $\es = \ew^2$. When $\es$ is below the threshold, additive noise prevails; above the threshold, Poisson noise dominates.

In Appendix~\ref{sec:upperBounds} we prove that the capacity $\capERC(\es,\ew)$ of the photon gas model is upper bounded by
\begin{align}\label{eq:capUB}
    \capERC(\es,\ew) &\leq \min (\capERC_\text{G}(\es,\ew),\capERC_\text{P}(\es)),
\end{align}
where $\capERC_\text{G}$ and $\capERC_\text{P}$ are respectively given by
\begin{align}
    &\capERC_\text{G}(\es,\ew) = H_\text{Geom}(\es+\ew) - H_\text{Geom}(\ew),\notag \\
    &\capERC_\text{P}(\es) = \log\Biggl(\biggl(1+\frac{\sqrt{2e}-1}{\sqrt{1+2\es}}\biggr)\frac{\bigl(\es+\frac{1}{2}\bigr)^{\es+\frac{1}{2}}}{\sqrt{e}\es^{\es}}\Biggr).     
\end{align}
Here $H_\text{Geom}(t)$ is the entropy of a geometric distribution with mean $t$, given by $H_\text{Geom}(t) = (1+t)\log(1+t) - t\log t$. Note that $\capERC_\text{P}(\es)$ does not depend on $\ew$.

Both functions $\capERC_\text{G}$ and $\capERC_\text{P}$ are monotonically increasing functions of $\es$. They have a crossing point, whose position we now determine under
the approximation $\ew \gg 1$ and $\es \gg \ew$. We can then use the asymptotic forms of the upper bounds to the capacity,
\begin{align}
    \capERC_\text{G}(\es,\ew) \simeq \log\biggl(\frac{\es}{\ew}\biggr)
    \simeq \frac{1}{2}\log(\es) \simeq \capERC_\text{P}(\es),
\end{align}
and we obtain again the expression
    $\frac{\es^2}{\ew^2} \simeq \es$,
previously derived by reasoning in terms of noise variance.

The threshold can be written in terms of the signal-to-noise ratio of the underlying classical channel,
\begin{align}\label{eq:thresholdSNR}
  \snr^* = \frac{E_s}{\sigma^2} = \frac{\es h\nu}{\ew h\nu} \simeq \ew \simeq \frac{k T}{h\nu} \simeq \frac{6\cdot 10^{12}}{\nu},
\end{align}
where $T = \units{290}{K}$ and $\nu$ is the frequency (in Hertz). In decibels, $\snr^*\text{(dB)} \simeq 37.8 - 10\log_{10}\nu$, where the frequency is in GHz.

Next to the upper bounds, we derive in Appendix~\ref{sec:lowerBound} a lower bound to the capacity. Its value is
\begin{align}
     \capERC_\text{LB} &= H_\text{Geom}(\es+\ew) - \dfrac{\ew}{\es+\ew}H_\text{Geom}(\ew) \notag\\ & \quad  - \dfrac{\es}{2(\es+\ew)}\Biggl(\log 2\pi e + \log\Bigl(\ew(1+\ew)+\tfrac{1}{12}\Bigr)\times  \notag\\ &\qquad \times e^{\frac{\ew(1+\ew)+\frac{1}{12}}{\es+\ew}}\Gamma\biggr(0,\frac{\ew(1+\ew)+\frac{1}{12}}{\es+\ew}\biggr)\Biggr),
\end{align}
where $\Gamma(0,t)$ is an incomplete gamma function,
    $\Gamma(0,t) = \int_t^\infty u^{-1}e^{-u}\,du$.

\begin{figure}[htbp]
\centering
  \includegraphics[width=\columnwidth]{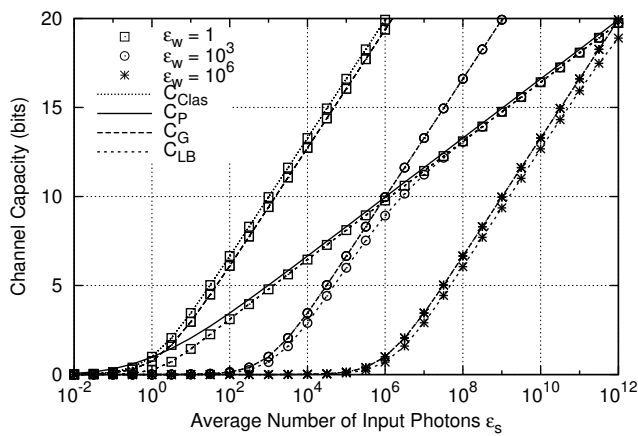}
  \caption{Bounds to the capacity for several values of $\ew$.}
  \label{fig:t20061026}
\end{figure}

The threshold can be seen in Fig.~\ref{fig:t20061026}, which depicts the upper and lower bounds to the capacity as a function of the input number of quanta $\es$ and for several values of $\ew$. The upper and lower bounds are close, differing by at most \units{1.1}{bits}. The looseness at low $\es$ is due to the pessimistic estimate of the conditional output entropy $H(\Zz|\Ss)$ (details are given in Appendix~\ref{sec:lowerBound}), which is smaller than the Gaussian approximation we have used. At high $\es$ the gap is likely due to the non-optimal input distribution, a gamma density with $\nu=1/2$, the density used in \cite{martinez07:spectralEfficiencyOpticalDirectDetection} on the bound the capacity of the discrete-time Poisson channel would likely close this gap. The capacity is closely given by the upper bound in Eq.~\eqref{eq:capUB}.

As proved in Appendix~\ref{sec:upperBoundBounds},
for finite values of $\es$, $\capERC_\text{G}(\es,\ew)$ is bounded by
\begin{align}\label{eq:corcapERCAEN-1}
    \log\Biggl(1 + \frac{\es}{\ew+1}\Biggr) < \capERC_\text{G}(\es,\ew) < \log\Biggl(1 + \frac{\es}{\ew}\Biggr).
\end{align}
The information rates over the photon gas model are strictly smaller than those of the classical limit with signal energy $E_s = \es h\nu$ and noise $\sigma^2 = \ew h\nu$. Nevertheless, the loss due to decoherence is negligible when $\es < \esStar$; here we assumed that the upper bound to the capacity is tight (see comments above and Sec.~\ref{sec:capacityAEN}).
On the other hand, above this energy level, the upper bound $\capERC_\text{P}$ becomes dominant, and eventually half of the achievable information rate is lost to decoherence.

A connection, worthwhile mentioning, can be made with non-coherent communications in Gaussian channels \cite{katz04:capacityAchievingNoncoherentAWGNChannels}, where one of the two signal quadratures is not used, and a similar change in slope in the capacity takes place. A similar limitation arises in phase-noise limited channels \cite{lapidoth02:phaseNoiseChannelsHighSNR}. As the threshold $\esStar$ is close to the point where existing digital communication systems using electromagnetic radiation suffer from the effects of phase noise, it would be interesting to verify which of the two models defines most accurately the effective channel capacity. Regarding this issue, note that the cost in information rates of maintaining the phase coherence between transmitter and receiver are usually ignored.


\subsection{Capacity over the Continuous Energy Model}
\label{sec:capacityAEN}
Under the approximation that the energy is continuous, Poisson noise vanishes, in the sense that the density of the random variable signal energy approaches a delta function,
    $p_{\Xx|\Ss}(\xk h\nu |\sk h\nu ) 
    \to \delta\bigl((\xk-\sk)h\nu\bigr)$.
In addition, the Bose-Einstein distribution collapses onto an exponential density.

The capacity of a channel with additive exponential noise was studied by Verd{\'u} \cite{verdu96:exponentialDistribution}. Applied to our channel model, we obtain the somewhat surprising
\begin{equation}
    \capAEN(E_s,\Ew) = \log\Biggl(1+\frac{E_s}{\Ew}\Biggr),
\end{equation}
as in the classical limit with Gaussian noise. Of course, this was to be expected since this model is a good description of the regime where $\capERC_\text{G}$ accurately gives the capacity, and the formula here follows from $\capERC_\text{G}$ as $\ew\to\infty$.

\section{Conclusions}


In this paper, we have compared two models for the transmission of information via electromagnetic radiation: the standard wave model with Gaussian noise and an alternative as a photon gas, where the radiation is an ensemble of photons over a set of Fourier modes. 
The second model represents a scenario with decoherence in the form of absence of phase reference between transmitter and receiver.
We have limited our analysis of decoherence to coherent states (in a semiclassical formulation) and do not consider quantum states such as squeezed states or Fock states, neither the use of entanglement. 

Even though decoherence makes the von Neumann entropy of the source radiation states increase \cite{caves94:quantumLimitsBosonic}, it does not necessarily lead to an information rate loss, even though the quadrature components of the field are not used separately. Essentially, the entropy of the received signal is determined by that of thermal radiation, if the signal energy is below a threshold.
A simple approximation for this threshold has been given. Below this threshold, there is no information loss caused by decoherence; above it, up to half of the chanel capacity is lost.

Capacity-achieving systems at radio and microwave frequencies operate below the threshold; optical systems are known to suffer from phase noise \cite{caves94:quantumLimitsBosonic}.
It would be interesting to find a practical example whose information rates, including the cost of maintaining coherence, would exceed that of the photon gas model. 
We conjecture that the information rates of practical communication systems are bounded by the capacity of the photon gas model and may be thus significantly smaller than those obtained from a Gaussian model. 

Finally, we mention that the photon gas model is somewhat close to a representation of classical matter as a set of particles, and that the results presented in this paper may be of help in exploring the quantum-classical border for radiation.
Differently from other examples of decoherence \cite{zurek03:decoherenceTransitionQuantumClassical}, the time scales at which einselection (in a quantum formulation of our semiclassical analysis) takes place are easily controllable, and therefore in principle observable, as opposed to decoherence in matter.

\appendix
\section{Upper Bounds}
\label{sec:upperBounds}
For any input $\pd_{\Ss}(\sss)$ the mutual information satisfies
\begin{align}
    \mInf{\Ss}{\Zz} &= H(\Zz) - H(\Zz|\Ss) \\
    &\leq H_\text{Geom}(\es+\ew) - H\bigl(\Xx(\Ss)+\Ww|\Ss\bigr),
\end{align}
as the geometric distribution has the highest entropy under the given constraints \cite{cover91:informationTheory}. Then,
\begin{align}
    H\bigl(\Xx(\Ss)+\Ww|\Ss\bigr) \geq H(\Ww|\Ss) = H(\Ww),
\end{align}
because the entropy of a sum of two independent random variables is at least as large as than the entropy of each of them (Exercise~18 of Chapter~2 of \cite{cover91:informationTheory}) and $\Ww$ and $\Ss$ are independent. Therefore,
\begin{align}
    \mInf{\Ss}{\Zz} 
    &\leq H_\text{Geom}(\es+\ew) - H_\text{Geom}(\ew).
\end{align}
As this holds for all inputs the upper bound $\capERC_\text{G}$ follows.

The variables $\Ss$, $\Xx(\Ss)$, and $\Zz(\Xx)$ form a Markov chain in this order,
$\Ss\to \Xx(\Ss) \to \Zz = \Xx(\Ss)+\Ww$, so that an application of the data processing inequality \cite{cover91:informationTheory} yields
\begin{equation}
    I\bigl(\Ss;\Zz\bigr) \leq I\bigl(\Ss;\Xx(\Ss)\bigr),
\end{equation}
that is the mutual information achievable in the discrete-time Poisson channel; a good upper bound to the capacity of the latter was given in \cite{martinez07:spectralEfficiencyOpticalDirectDetection}.

\section{Lower Bound}
\label{sec:lowerBound}
Our lower bound is derived from the mutual information achievable by a specific input with density
\begin{equation}\label{eq:distX-AEQ}
    \pd_{\Ss}(\sss) =
    \dfrac{\es}{(\es+\ew)^2}e^{-\frac{\sss}{\es+\ew}}+\dfrac{\ew}{\es+\ew}\delta(\sss), \quad \sss \geq 0.
\end{equation}
It is easy to prove that the channel output induced by this input achieves the largest output entropy; a similar proof can be found in \cite{verdu96:exponentialDistribution} for the exponential noise channel.
As a particular case we recover the exponential input, which maximizes the output entropy of a discrete-time Poisson channel \cite{gordon62:quantumEffectsCommunicationSystems}.

By construction, the output is Bose-Einstein (geometric) with mean $\es+\ew$ and the output entropy $H(\Zz)$ is therefore given by
    $H(\Zz) = H_\text{Geom}(\es+\ew)$. We compute the mutual information with this input as
    $H(\Zz) - H(\Zz|\Ss)$.

We estimate the conditional entropy as
\begin{align}
    H(\Zz|\Ss) &= \int_0^\infty H(\Zz|\sss)\,\pd_{\Ss}(\sss)\,d\sss.
\end{align}
We obtain a term $\frac{\ew}{\es+\ew}H(\Zz|\sss=0)$, which can be computed as
    $H(\Zz|\sss=0) = H_\text{Geom}(\ew)$.
A second summand is upper bounded by the differential entropy of a Gaussian random variable (see Theorem~9.7.1 of \cite{cover91:informationTheory}),
\begin{align}
    H(\Zz|\sss) &\leq \frac{1}{2}\log 2\pi e\Bigl(\Var (\Zz|\sss)+\tfrac{1}{12}\Bigr)\\
    &= \frac{1}{2}\log 2\pi e \Bigl(\sss+\ew(1+\ew)+\tfrac{1}{12}\Bigr).
\end{align}
The desired expression follows from carrying out the integration and using the definition of the incomplete gamma function.

\section{Bound Estimates}
\label{sec:upperBoundBounds}
First, we prove the strict inequality
\begin{align}
    \log&(\es+\ew)-\log\ew > \capERC_\text{G}(\es,\ew),
\end{align}
for all values of $\es > 0$, $\ew \geq 0$.
Using the definition of $\capERC_\text{G}$, we rewrite this expression as
\begin{align}
    (1+\es+\ew)&\log\frac{\es+\ew}{1+\es+\ew} > (1+\ew)\log\frac{\ew}{1+\ew}.
\end{align}
Proving this is equivalent to proving that the function
    $f(t) = (1+t)\log\frac{t}{1+t}$
is monotonically increasing for $t > 0$. It is indeed so since its first derivative $f'(t)$ is
\begin{align}
    f'(t) 
    &= \frac{1}{t} -\log\biggl(1+\frac{1}{t}\biggr),
\end{align}
which is positive since $\log(1+t') < t'$ for positive $t'$.

We now move on to prove 
\begin{align}
    \capERC_\text{G}(\es,\ew)  > \log(\es+\ew+1)-\log(\ew+1).
\end{align}
From the definition of $\capERC_\text{G}$, and after cancelling common terms, we rewrite the condition as
\begin{align}
    (\es+\ew)\log\frac{1+\es+\ew}{\es+\ew} > \ew\log\frac{1+\ew}{\ew}.
\end{align}
This equation is true because the function 
    $f(t) = t\log\bigl(1+\frac{1}{t}\bigr)$
is monotonically increasing for $t > 0$, since it monotonically approaches the number $e$ from below.


\end{document}